\documentclass[10pt, preprint2]{aastex}

\begin{document}

\title{More Evidence for an Oscillation Superimposed on the Hubble Flow}



\author{M.B. Bell\altaffilmark{1} and S.P. Comeau\altaffilmark{1}}

\altaffiltext{1}{National Research Council of Canada, 100 Sussex Drive, Ottawa,
ON, Canada K1A 0R6; morley.bell@nrc-cnrc.gc.ca}

 \begin{abstract}

In a recent investigation evidence was presented for a low-level sinusoidal oscillation superimposed on top of the Hubble flow. This oscillation was in V$_{CMB}$, in a sample of type Ia Supernovae sources with accurate distances, and it was found to have a wavelength close to 40 Mpc. It became easily visible after the removal of several previously identified discrete velocity components. Its amplitude like that of the Hubble velocity showed an increase with distance, as would be expected for a constant-amplitude space oscillation. Here we report that this oscillation is also present in distance clumping in these sources, with the same wavelength, but in phase quadrature. The discrete velocity components do not play a role in detecting the distance clumping wavelength. Assuming that time proceeds from high cosmological redshift to low, the blue-shifted velocity peaks, which represent the contraction stage of the velocity oscillation, then lead the density peaks. With the discrete velocity components removed we also find evidence for at least one other, weaker velocity oscillation. It is found to have a wavelength similar to one reported in density clumping by previous investigators. In those cases the source samples were much larger.

\end{abstract}





\section{Introduction}

Recently, using the SneIa data from \citet{fre01}, where special precautions were taken to insure that the source distances were accurate, we have shown that there is evidence for a low-level oscillation superimposed on the Hubble flow that has a wavelength near 40 Mpc \citep{bel13}. This result was obtained after determining and removing discrete velocity components of the form discussed by \citet{tif96,tif97}, by us \citep{bel03a,bel03b}, and by \citet{dav05a,dav05b,dav05c}. Note that that determining the discrete velocity components is only possible when the source distances are accurately known. Removal of these components resulted in a significant reduction in the RMS deviation in V$_{CMB}$ from 780 to 166 km s$^{-1}$.
Here we report that, in addition to the previously reported velocity oscillation, there is also evidence for density clumping visible in the distance distribution of SneIa galaxies observed by \citet{fre01}. It too has a wavelength near 40 Mpc. This is not unexpected as \citet{mor90} has pointed out that a density oscillation can be produced by a velocity oscillation superimposed on the Hubble flow. Unlike the velocity oscillation, the density clumping is in distance and the discrete velocity components do not play a role in its detection. We also search here for other velocity periods present in the SNeIa data after the discrete components are removed. We show in Section 4 that not only does at least one other, weaker velocity period become visible after the discrete components are removed, it has a period similar to that found in density clumping by other investigators.
Throughout this paper the term \em contraction stage \em refers to the blue-shifted portion of the velocity oscillation.

\begin{figure}[h,t]
\hspace{-0.8cm}
\vspace{0.0cm}
\epsscale{0.9}
\includegraphics[width=9cm]{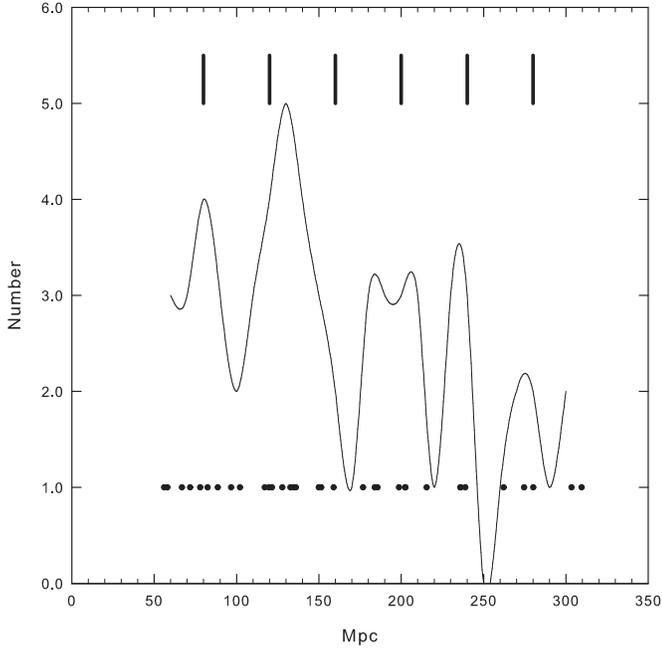}

\caption{\scriptsize{Distances of SneIa sources from \citet{fre01} are plotted at level 1. The curve shows a smoothed distribution of the data using a bin width of 20 Mpc sampled every 10 Mpc. This curve shows several peaks that agree reasonably well with the 40 Mpc period indicated by the vertical bars near the top. \label{fig1}}}
\end{figure}

\begin{figure}[h,t]
\hspace{-0.8cm}
\vspace{-1.5cm}
\epsscale{0.9}
\includegraphics[width=9cm]{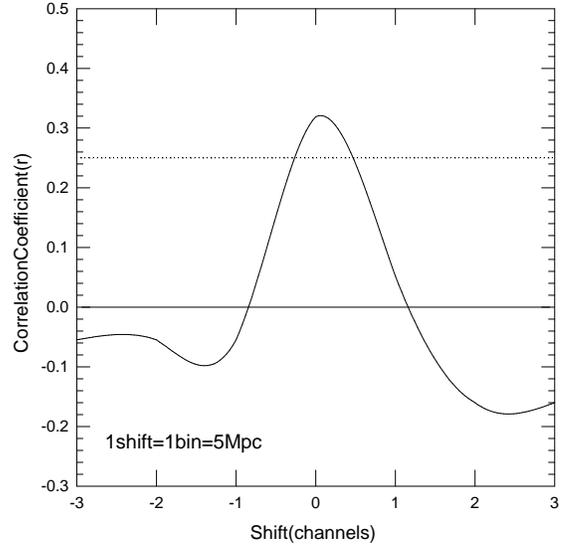}
\caption{{Correlation coefficient (r) obtained as described in the text and plotted as a function of phase of the comparison 40 Mpc distribution. Zero shift corresponds to the case where the phase of the comparison distribution was chosen so peaks were at 40, 80, etc., Mpc.  \label{fig2}}}
\end{figure}

\begin{figure}[h,t]
\hspace{-1.0cm}
\vspace{-2.0cm}
\epsscale{0.9}
\includegraphics[width=9cm]{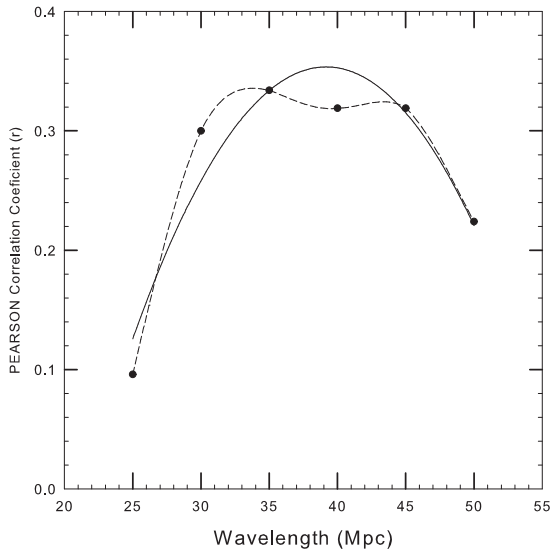}
\caption{\scriptsize{Plot of correlation coefficient vs wavelength of the density ripple. Solid curve represents the Gaussian fit to the correlation data. \label{fig3}}}
\end{figure}

\section{Evidence for Density Clumping.}

In Fig 1 the distances of the SneIa sources are plotted at level 1. The curve in Fig 1 shows a smoothed distribution of the data obtained by taking a running source count using a bin width of 20 Mpc and moving the bin in 10 Mpc steps. It shows at least five peaks that agree reasonably well with the vertical bars near the top of the figure that have a 40 Mpc spacing. This preliminary look then suggests that there may well be density clumping in the source distribution that also has a wavelength close to 40 Mpc.

To test this further, and to determine if this apparent clumping is significant, we carried out a correlation analysis between, (a) the distance distribution of sources as plotted at level 1 in Fig 1 and, (b) distributions where the source spacing was purely periodic. In this latter case the different source spacings, or wavelengths, tried ranged from 35 to 55 Mpc, in 5 Mpc steps.

To do this we first binned each of the groups into 60 equal bins between 0 and 300 Mpc, and counted the number of sources in each bin. We then determined the best correlation coefficient between the data distribution and each of the purely periodic distributions by varying the phase of each of the latter. The result for the 40 Mpc wave is shown in Fig 2. Here the zero shift position corresponds to a phase with peaks at 40, 80, etc., Mpc. The correlation coefficient drops quickly when the 40 Mpc distribution is shifted relative to distribution (a). The best-fit phase is $+0.18\pm0.2$ shifts, or $0.9\pm1$ Mpc (1 shift = 1 bin = 5 Mpc), giving density peaks located at 40.9, 80.9, 120.9, etc., $\pm1$ Mpc. Since the velocity wave had redshift peaks at 29.7, 69.7, 109.7, etc., Mpc, this indicates that the density clumping is $11.2\pm1.3$ Mpc, or $100.8\pm12$ degrees out of phase with the velocity peaks. Thus, within the error, the density and velocity waves are in phase quadrature.

As can be seen in Fig 2, a correlation coefficient of 0.31 was obtained for distributions with wavelengths near 40 Mpc. For 60 bins, a correlation coefficient of 0.25, indicated by the horizontal dotted line in Fig 2, is considered significant \citep{whe68}.

\begin{table}

\caption[]{Parameters of Velocity and Density Waves. \label{tab-1}}
$$
\begin{tabular}[]{cc}

\hline

Parameter & Value \\

\hline

$\lambda$(Velocity wave)    & 39.6$\pm$0.1 Mpc \\ 
Velocity peaks   & 29.7, 69.7,109.7,..  Mpc \\ 


$\lambda$(Density wave)  &  39.2$\pm$1 Mpc  \\ 

Density peaks  &  40.9, 80.9, 120.9,.. Mpc  \\ 
\hline

\end{tabular}
$$

\end{table}

\begin{figure}[h,t]
\hspace{-0.8cm}
\vspace{0.0cm}
\epsscale{0.9}
\includegraphics[width=9cm]{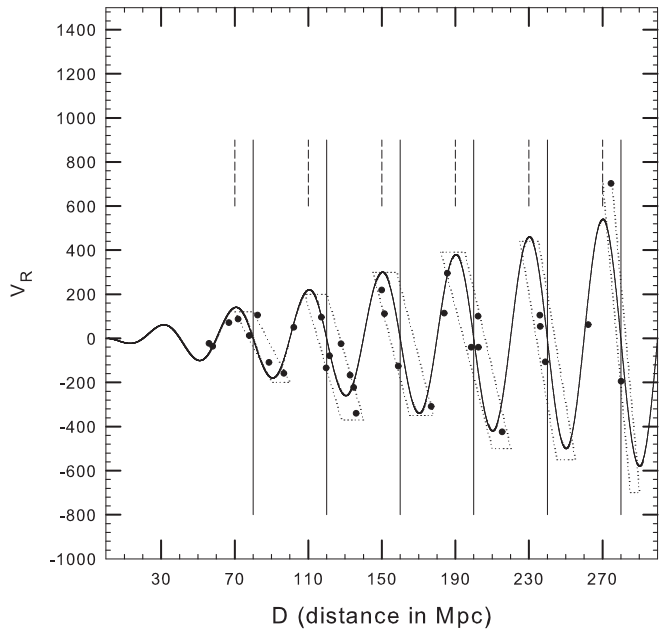}
\caption{\scriptsize{Plot of residual velocity versus distance for SNeIa galaxies after removal of intrinsic redshifts and Hubble slope of 57.9 km s$^{-1}$ Mpc$^{-1}$. Velocity peaks are located at 30, 70, 110, 150, 190, 230 and 270 Mpc. Regions of high source density are indicated by the dotted rectangles and are centered near 80, 120, 160, 200, 240 and 280 Mpc, as indicated by the solid vertical lines, and are in phase quadrature with the velocity peaks indicated by the dashed vertical lines. \label{fig4}}}
\end{figure}

The highest correlation coefficient for each of the periodic distributions (obtained by optimizing the phase) was then plotted as a function of wavelength and is shown here in Fig 3. This plot shows a broad feature centered near 40 Mpc. A Gaussian fit to this feature gave a best-fit density wavelength of 39.2$\pm1$ Mpc. Parameters for the velocity and the best-fit density wave are included in Table 1.

The relative phases of the velocity and density waves can also be seen in Fig 4 where the residual velocity, V$_{R}$, of the SneIa galaxies studied by \citet{fre01} is plotted versus distance. V$_{R}$ was obtained by removing a Hubble slope of 57.9 km s$^{-1}$ Mpc$^{-1}$ from the Hubble velocities, V$_{H}$, obtained for the SneIa galaxies. The V$_{H}$ values are listed in Table 1 of \citet{bel13}, and were obtained after removing discrete velocity components as discussed previously \citep{bel13}.

In Fig 4 the vertical dashed lines indicate the peaks of the velocity wave shown by the sinusoidal curve. The vertical solid lines have been drawn at the zero-level crossings of the negative-slope portion of the sine wave. The dashed and solid lines are then 90 degrees apart in phase. In Fig 4 it is also apparent that there is a higher density of sources located in the region of the negative slopes of the sinusoidal curve (indicated by the dotted rectangles) than in the regions of positive slope. In fact, for $D > 70$ Mpc, only 5 of the sources are located on the positive slopes. This indicates clearly that the density peaks in the source distribution are centered near the zero-crossings of the negative-slope portion of the sinusoid, which is 90 degrees from the peaks of the velocity wave, confirming our previous result that the velocity and density waves are in phase quadrature.

 
To be able to find these relationships in the data by chance seems very unlikely. It is also worth noting that the density clumping has nothing to do with whether or not there are discrete components present in the velocity data, since it examines only the distances provided by \citet{fre01}. Although the clumping in the density wave is not as obvious as the ripple seen in the velocity wave, it is rather remarkable that they both share the same period and are in phase quadrature. If it is assumed that time proceeds from high cosmological redshift to low, the blueshifted velocity peaks (contraction stage) then lead the density peaks.

\section{Are there other Velocity Periods present?}

When the solid curve in Fig 4 was removed from the data, five of the sources (SN 1995ak, SN 1994Q, SN 1993ac, SN 1992bh, and SN 1992ae) were seen to have discrepant velocities in that they all had velocities that were +200 to +400 km above the mean. The fact that this component was always positive suggests that the distances of these sources may still contain an error that might then have resulted in an incorrect identification of its intrinsic component. For this reason these sources were not included in the following analysis.


In order to investigate whether or not other periods were present in the velocity data we took the following approach. A program was written that found the best-fit, constant amplitude sine waves in the data in Fig 4 (fitting amplitude and phase) for a range of wavelengths from 25 to 250 Mpc, in 5 Mpc intervals. When the best-fit sinusoid at each wavelength was determined, it was removed from the data and the new RMS value obtained was tabulated. The resulting RMS values are plotted here versus wavelength in Fig 5. The horizontal line represents the RMS value obtained for the raw data in Fig 4. In Fig 5 the vertical scale has been inverted so the relative strengths of the features indicate the relative strengths of any sinusoids present in the distribution. As expected, the best fit is obtained for a wavelength near 40 Mpc, which has an off-scale RMS value of 96. However, as can be seen in Fig 5, there is another peak visible, located at a wavelength near 87 Mpc. There is also a broad peak located near 187 Mpc but its weak strength may make its validity questionable. However, in Fig 6 the curve in Fig 5 has been re-plotted on a logarithmic scale where the three peaks show a log-periodic nature, presumably resulting from the fact that the wavelength of the peaks in Fig 5 increases by a factor close to two. Because of this apparent harmonic relationship we were concerned that the weaker periods might be spurious, and simply produced by the analysis. To check this we repeated the analysis replacing the data with a pure sinusoid of wavelength 40 Mpc. However, in this case no evidence for harmonically related features was found. We therefore concluded that the weaker wavelengths are likely to be real.

\section{Discussion}

After identifying and removing small discrete velocity components from the radial velocities of SNeIa galaxies we have previously found evidence for a sinusoidal oscillation in V$_{CMB}$ on top of the otherwise linear Hubble plot. It has a relatively short wavelength of 39.6 Mpc (period = 1.3$\times10^{8}$ yrs, frequency = 2.4 x 10$^{-16}$ Hz).
Here we have found that it has an associated density clumping that has an identical wavelength, but with density peaks that follow the blue-shifted velocity peaks (contraction stage), in phase quadrature.

\begin{figure}[h,t]

\hspace{-0.8cm}
\vspace{-1.0cm}
\epsscale{0.9}
\includegraphics[width=9cm]{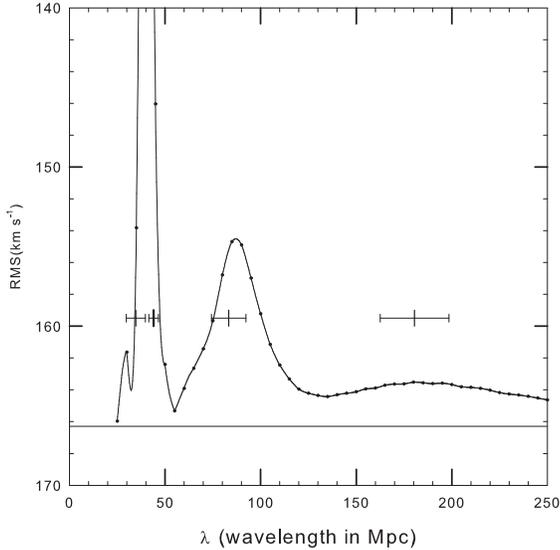}
\caption{\scriptsize{Plot of RMS deviation in V$_{\rm R}$ for data in Fig 1 after removal of best-fit sine waves of various wavelengths between 25 and 250 Mpc. The vertical axis has been inverted so the plot shows the relative strengths of each component. The vertical bars plotted at wavelengths near 34.7 and 83.3 Mpc, at RMS level 159.5, show galaxy distribution scales reported by \citet{moh92}. The vertical bar at 44$\pm2.5$ Mpc indicates the wavelength of a density clumping reported by \citet{har08}. The vertical bar at 180.5 Mpc indicates the location of the density clumping found independently by several other groups at $\sim130h^{-1}$ Mpc (see text for further discussion). \label{fig5}}}

\end{figure}
 
\begin{figure}[h,t]
\hspace{-0.8cm}
\vspace{-1.0cm}
\epsscale{0.9}
\includegraphics[width=9cm]{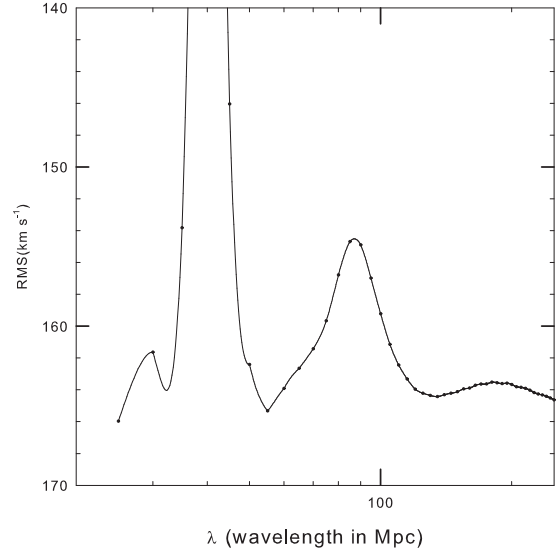}

\caption{\scriptsize{Same as fig 4 plotted on a logarithmic scale to show the log-periodic nature of the different velocity periods. \label{fig6}}}
\end{figure}

The characteristics of the 39.6 Mpc velocity and density oscillations seem to imply that as space expands, it also oscillates, and that this oscillation, at least over the epoch we have accessed, has a constant amplitude.
Although the sources in each clump will be located in a shell centered on the Earth, this does not imply that there are alternating concentric high- and low-density shells centered on our location in the Universe. It simply means that because of the look-back time we are observing the Universe at various stages of this high- and low-density contraction and expansion cycle.
To be able to find a ripple in velocity and density with these exact characteristics by chance would seem to be very unlikely, and it implies that this result may well have potentially significant implications for cosmology.

Clumping in the distribution of galaxies has been reported by several groups (see e.g., \citet{bro90,huc90,row90,moh92,lan96}). \citet{moh92} found three typical clumping scales in the distribution of galaxies in this range. These were at 16$h^{-1}\pm1$ Mpc ($h$ = H$_{\rm o}$/100)\footnote{The density-clumping scales can be converted to distances by assuming a value for the Hubble constant. Since these investigators used data containing the discrete velocity components, the appropriate value for the Hubble constant should be close to $h$ = 0.72 (the value found in the Hubble Key Project \citep{fre01} where velocities that contained the discrete components were also used.)}, 25$h^{-1}\pm2$ Mpc, and 60$h^{-1}\pm4$ Mpc (see figure 7 of \citet{moh92}).


When the density clumping scale 25$h^{-1}$ is converted to distance, it corresponds to a wavelength of 34.7 Mpc which is in reasonably good agreement with the dominant velocity wavelength of 39.6 Mpc found here. Similarly, the 60$h^{-1}\pm4$ value converts to 83.3$\pm5$ Mpc, which is in good agreement with the 87.5 Mpc value found above in velocity in Fig 4.

A larger scale size of $\sim130h^{-1}$ Mpc was detected in the deep pencil-beam survey of galaxies \citep{bro90}, the deep redshift survey of Abell clusters \citep{huc90} and the IRAS QDOT sample \citep{row90}. This converts to 180.5 Mpc which, although very weak, is also visible in Fig 4.
 
Our strongest observed ripple, with a wavelength near 40 Mpc, is also similar to one of the density clumping wavelengths reported previously by \citet{har08}. Their distance spacing of (31.7$\pm1.8)$h$^{-1}$ corresponds to 44$\pm2.5$ Mpc, and has also been plotted in Fig 4. Since the results of \citet{moh92} at 34.7 and \citet{har08} represent independent results, they can be averaged. This results in the value 39.35 Mpc, which is almost identical to our velocity value of 39.6 Mpc. Thus the density clumping wavelengths found by others, which correspond to 34.7 Mpc, 44 Mpc, 83.3 Mpc, and 180.5 Mpc, agree well with our velocity wavelengths of 39.6 Mpc, 87 Mpc and 187 Mpc. This is especially true when it is taken into account that many of the density clumping results were obtained using redshifts instead of accurate distances. This means that there would be some smearing present because the redshifts from which the distance are obtained will still contain the velocity ripple as well as a small portion of the discrete components, even though most of this is accounted for by the difference between the Hubble slopes used (our 58 versus their 72 km s$^{-1}$ Mpc$^{-1}$).

We conclude that the $velocity$ wavelengths we report here, and previously \citep{bel13}, appear to have been confirmed by the results found for $density$ wavelengths reported here by us, and previously by \citet{moh92,bro90,huc90,row90}. We can also conclude that this confirmation of our velocity wavelengths by the work on density clumping carried out by previous investigators offers an equally strong argument in favor of the reality of the discrete components, since the velocity fluctuations in the Hubble flow reported both here, and previously for the SNeIa galaxies, could not be seen before the unique, discrete components were removed from their radial velocities. This is also a confirmation of both Tifft's work \citep{tif96,tif97} and our work on discrete velocities in other galaxy samples \citep{bel03a,bel03b,rus03}.

Although it might seem at first glance that the ripple amplitude increases with distance, this is unlikely to be the case. This apparent increase with distance is exactly what is expected for constant ripple amplitude if, like the Hubble flow, it is space itself that is oscillating.

The galaxy distribution scales of $25h^{-1}$ (34.7) Mpc and $60h^{-1}$ (83.3) Mpc have been attributed by \cite{moh92} to the separations between clusters and superclusters respectively. However, our results suggest that this interpretation may need to be looked at more closely.

The various galaxy distribution studies listed above have since spawned several papers attempting to explain the density oscillations \citep{mor90,mor91,hil91,bus94,sal96,ein97,hir10}. Although it is beyond the scope of this paper, the models discussed in some of these papers may also need to be looked at again in light of our results.

\section{Conclusions} 

We have previously identified discrete velocity components in the radial velocities of the 36 Type Ia Supernovae galaxies studied in the Hubble Key Project. We have shown that when these intrinsic components are removed from the redshifts of the SNeIa galaxies, the Hubble constant obtained is close to 58 km s$^{-1}$ Mpc$^{-1}$ and that there is evidence for a low-level sinusoidal oscillation superimposed on the Hubble flow. It is isotropic in nature and has a wavelength of 39.6 Mpc. Here we find that this same oscillation is present in density clumping. The two are in phase quadrature, with the blue shifted velocity peaks leading the density peaks. The density and velocity fluctuations have wavelengths of 39.2 and 39.6 Mpc respectively. This is the first time that both velocity and density fluctuations have been detected in the same data and, consequently, the first time the relative phases of the velocity and density waves could be determined.
It seems very unlikely that a systematic effect could have somehow produced the observed oscillation in the Hubble flow, especially since the distance clumping, which is independent of any assumption of discrete velocities, shows an identical wavelength structure. We have also found evidence for at least one other velocity oscillation. It has a wavelength near 87 Mpc. Although much weaker, a third velocity oscillation may also be present at a wavelength near the value reported by Broadhurst (1990). All of the observed velocity wavelengths are found to be similar to those found previously by other investigators to be present in density clumping. This appears to confirm our findings, which in turn confirms the validity of the discrete components because they have to be removed before the velocity ripples in the Hubble flow become visible.

\section{Acknowledgements}

We thank Donald R. McDiarmid for helpful comments.


\end{document}